\documentclass[%12pt,
amsmath, aps, prl, twocolumn,showpacs
]{revtex4}
\usepackage{bm}
\usepackage{graphicx}
\usepackage{amssymb}

\def \d{\partial}
\newcommand{\dd}{\partial}
\def \bv{{\bf v}}

\def \br{{\bf r}}

\def \bB{{\bf B}}

\def \bR{{\bf R}}
\def \br{{\bf r}}

\def \bk{{\bf k}}

\def \bphi{\boldsymbol{\phi}}

\def \bomega{\boldsymbol{\omega}}
\def \bOmega{\boldsymbol{\Omega}}

\def\bfeta{\boldsymbol{\eta}}
\def \eps{\varepsilon}

\def \varkappa{\kappa}

\begin{document}

\title{Stationary solution for quasi-homogeneous small-scale magnetic field advected by non-Gaussian turbulent flow}

\author{ A.S. Il'yn$^{1,2}$, A.V. Kopyev$^1$, V.A. Sirota$^{1}$, and K.P.
Zybin$^{1,2}$\thanks{%
Electronic addresses: asil72@mail.ru,  kopyev@lpi.ru, sirota@lpi.ru,
zybin@lpi.ru}}
\affiliation{$^1$ P.N.Lebedev Physical Institute of RAS, 119991, Leninskij pr.53, Moscow,
Russia \\
$^2$ National Research University Higher School of Economics, 101000,
Myasnitskaya 20, Moscow, Russia}

%\pacs{47.10.+g}{General theory in fluid dynamics}
%\pacs{47.27.tb}{Turbulent diffusion}
%\pacs{47.65.-d}{Magnetohydrodynamics and electrohydrodynamics }
\pacs{47.10.+g,  47.27.tb, 47.65.-d}

\begin{abstract}
We consider fluctuations of magnetic field excited by external force and advected by isotropic turbulent flow.
% advection of forced magnetic field by a turbulent flow in  the viscous range of isotropic turbulence.
It appears that non-Gaussian velocity gradient statistics and finite region of pumping force provide the existence of stationary  solution.
% convergence of the
%second-order correlator.
The mean-square magnetic field is  calculated for arbitrary velocity gradient statistics. An estimate for possible feedback of magnetic field on velocity  shows that, for wide range of parameters, stationarity without feedback would take place even in the case of intensive pumping of magnetic field.
\end{abstract}

\maketitle

\section{Introduction}
Despite numerous papers and many impressive results, %classical problems of
turbulence remains one of unsolved fundamental problems in physics.
In particular, in fully developed stationary turbulence anomalous scaling is still unexplained: it is observed in many experiments \cite{anom-scaling-exp} and simulations \cite{anom-scaling-sim}, and described by phenomenological theories (e.g., multifractal theory, \cite{multifrac25}) but no physical understanding of the effect has been achieved.

Fully developed turbulence is very complicated object. %, and the equations describing it are difficult to study.
To gain some experience in similar but easier problems, it is useful to consider statistically stationary configurations in other processes. In particular, advection of passive scalar and vector fields by a turbulent flow %with given characteristics
is the natural training ground.

The passive field advection problem has also its own significance \cite{FGV}. Scalar advection is demanded in chemical and biological applications, and magnetic field advection   by a turbulent flow is conventionally considered as an origin of magnetic field in different magneto-hydrodynamic systems, e.g., planets, stars, media, galaxies and galaxy clusters \cite{Moffat,Parker}. One distinguishes two types of magnetic field and, accordingly,  two types of turbulent dynamo that is responsible for its production \cite{Rincon,Brandenburg12}. The first type is large-scale fields with characteristic scales of the order of the size of the host object. Individual features of the object are essential in this case. The second type is small-scale magnetic fields with scales much smaller than the size of the object and than the scale at which turbulence is generated; this corresponds to inertial or viscid scale ranges of turbulence. In \cite{Kulsrud92} it was shown that this type may be even more efficient, since the small-scale dynamo provides faster growth of magnetic field.
Probably, both types are present in astrophysical objects.

In what follows we restrict our consideration by statistically stationary small-scale fields driven by a turbulent flow.
The characteristics of the flow are assumed to be known.

The stationary  solution for homogeneous passive scalar field pumped by a random external force and advected by a turbulent flow was found in \cite{Chertkov95,BF} for a viscid flow, and in \cite{GawKup,Ant99} for the inertial range. For  non-divergent vector (magnetic) field, the advection in the inertial range of turbulence was considered in \cite{Vergassola96}.
 In the viscous range, stationary solution has not been considered up to recent time because in \cite{Kazantsev, Chertkov} it was shown that small-scale turbulent dynamo leads to infinite increase of the advected magnetic field even in the absence of  pumping force. So, it seemed evident that stationary magnetic field configuration in the viscous range of turbulence is only possible if the advected field is strong enough to affect the velocity dynamics. Thus, the feedback of magnetic field on the advecting flow  seemed to be necessary to provide statistical stationarity.

However, the infinite growth of magnetic field in \cite{Chertkov} was obtained under several assumptions: homogeneous initial conditions for magnetic field, isotropy and Gaussianity of velocity gradients. It appears that these assumptions, being representative in the case of scalar field, do not present the general case for the advected vector. In \cite{ZMRS} it was shown that in some special step-like velocity field, a finite-size initial magnetic blob would decay; in \cite{epl} the same result was obtained for arbitrary non-Gaussian velocity gradients. So, infinite growth of advected magnetic field is only the property of Gaussian statistics and/or  homogeneous initial conditions; to the contrary, in general case the stationary solution is possible.

In \cite{PREnew} we considered the evolution of magnetic field pumped by inhomogeneous driving force and advected by a viscid turbulent flow, and found a stationary solution for spatial correlators. However, the range of the smallest scales (smaller than the diffusivity scale) was not investigated. So, it remained unknown whether   the stationary one-point correlator exists or not. This is not evident: the well-known example of spatially homogeneous pumped passive scalar field  in the case of zero diffusivity provides the situation in which stationary two-point correlator exists for any finite distance between the points, but the one-point correlator diverges as $t \to \infty  $ \cite{Chertkov95}. So, the scalar field remains statistically non-stationary, although the scale of non-stationarity decreases as a function of time; the existence of finite stationary correlator for any finite distance still does not guarantee the existence of fully statistically stationary solution.

In this paper we examine the mean-square magnetic field; we show that in presence of random inhomogeneous pumping force, statistically stationary solution is possible: one-point correlator converges, and the convergence of the two-point correlator is uniform.

We consider the viscous range of incompressible turbulence, assuming  velocity gradients statistics known. The advected magnetic field is assumed to have diffusive scale $r_d$, the pumping correlation scale $l$, and the region of pumping restricted by the scale $L$ such that
\begin{equation} \label{inequalities}
  r_d \ll l \ll L \ll r_{\eta}
\end{equation}
where $r_{\eta}$ is the Kolmogorov viscous scale. We calculate the mean-square magnetic field and prove that it is finite for general non-Gaussian field and finite $L$. We also estimate the conditions  that provide a feedback of magnetic field on velocity statistics, and show that  stationary solution is possible without the feedback.

%\vspace{1cm}

\section{Problem statement}
Transport of magnetic flux density $\bB(\br,t)$ driven by a random force $\mathbf{\phi}(\br,t)$ in external random velocity field  $\bv (\br,t)$ obeys  the equation:
\begin{equation}\label{1}
\frac{\dd \mathbf{B}}{\dd t}+\bigl(\mathbf{v} \nabla \bigr)\mathbf{B}-\bigl(\mathbf{B}\nabla\bigr)\mathbf{v} =  \varkappa \Delta \mathbf{B} + {\bphi}
\end{equation}
Here $\varkappa$ is the magnetic diffusivity.

The last inequality in (\ref{inequalities}) means that the velocity field is smooth, and in quasi-Lagrangian frame \cite{Lvov} we arrive to the Batchelor approximation \cite{Batchelor},
\begin{equation} \label{v=Ar}
\bv = {\bf A} \br \ ,    %v_i = A_{ij} r_j \ ,
\end{equation}
the velocity gradient tensor ${\bf A}(t)$ is a random process with short correlation time, $\tau_c \ll t$.
The incompressibility condition results in ${\rm Tr} {\bf A}=0$.
%The Lyapunov indices  are the main characteristics of the tensor \cite{Oseledets}; in this paper we use the ordering  $\lambda_1<\lambda_2<\lambda_3$.
The diffusive scale $r_d$ is determined  %related to $\varkappa$
by $r_d = \sqrt{\varkappa /\lambda_3}$, where $\lambda_3$ is the maximal Lyapunov exponent \cite{Oseledets}.

The pumping force is assumed to be Gaussian. From (\ref{1}) it follows that $\phi$ must be solenoidal;
 the pair correlator of the Fourrier transform $\tilde{\bphi} (\bk,t)$  is chosen
\begin{equation}\label{E:f_corr_inhom}
\begin{array}{r} \displaystyle
\langle\tilde{\phi}_i(\mathbf{k},t)\tilde{\phi}_j(\mathbf{k}',t')\rangle=
\frac {4}{3\pi^2} \eps_B L^3 l^5 \mathrm{e}^{-\frac{1}{4}(\mathbf{k}+\mathbf{k}')^2 L^2 %}\mathrm{e}^{
-\frac{1}{4}(\mathbf{k}-\mathbf{k}')^2 l^2}
\\ \times
\bigl(k_j\,k'_i-\left(\mathbf{k}\,\mathbf{k}'\right)\delta_{ij}\bigr)\delta(t-t') \ ,
\end{array}
\end{equation}
where $\eps_B= \frac 1{4\pi} \langle  \bB (0) \cdot {\boldsymbol{\phi}}(0) \rangle$ is the pumping power,
\begin{equation}
 \label{phi-kvadrat}
 \left \langle \phi(\br,t)  \phi(\br,t') \right \rangle = 2\pi \eps_B \delta(t-t')
\end{equation}

Our aim is to calculate
\begin{equation} \label{beta-def}
\beta_0 =  \left \langle \bB (0,t)^2 \right \rangle _{\phi,A}
\end{equation}
where the average is taken over both the velocity gradient and the pumping force.

Substituting (\ref{v=Ar}) for $\bv$ in (\ref{1})  and solving the linear equation, we get
\begin{equation}\label{B(k)}
\begin{array}{r} \displaystyle
B_p({\bf 0},t)=  \int d\bk \int_0^t\mathrm{d} \tau W_{pj}(t,\tau) \tilde{\phi}_j\bigl(\mathbf{k}\mathbf{W}(t,\tau),t-\tau\bigr)
\\  \times
\mathrm{e}^{-\varkappa\,
\mathbf{k}\bigl(\int_0^{\tau}\mathbf{W}(t,\tau')\mathbf{W}^{T}(t,\tau')\mathrm{d} \tau'\bigr)\,\mathbf{k}^T}
\end{array}
\end{equation}
where ${\bf W}(t,\tau)$ is the evolution matrix that  satisfies
\begin{equation}\label{evolution_W}
\d {\bf W} / \d \tau = {\bf W}(t,\tau) {\bf A}(t-\tau)  \ , \qquad  {\bf W}(t,0)=1
\end{equation}
Since $\bf A$ is a random process, $\bf W$ is also random and its statistics is determined by the statistics of $\bf A$.

\section{ Averaging over $\mathbf \phi$} \ \
Substituting (\ref{B(k)}) into (\ref{beta-def}) and making use of (\ref{E:f_corr_inhom}), we get
\begin{equation} \label{beta-average}
 \beta_0 =\left \langle \int d \tau \Psi \left[ \tau \, , {\bf \bOmega}(t) \right] \right \rangle_A
\end{equation}
where
\begin{equation} \label{psi-polu}
\begin{array}{rl}
\Psi  \propto&  \varepsilon_B   \int d\bk d\bk'  \bk   \left(  \bOmega^2  -  \bOmega  {\rm Tr} \bOmega   \right) \bk'
 e^{ -\frac14 (\bk-\bk')\bOmega (\bk-\bk')^T l^2  } \\
 \\
\times& e^{  -\frac14 (\bk+\bk')\bOmega (\bk+\bk')^T L^2
-\varkappa \bk \int_0^{\tau} \bOmega d \tau' \bk -\varkappa \bk' \int_0^{\tau} \bOmega d \tau' \bk' } , \\
&  \ \  \bOmega = {\bf WW}^T
\end{array}
\end{equation}

We see that the 'viscous' terms in the exponent appear in  combinations
$ \frac 14 l^2 \bOmega + \varkappa  \int_0^{\tau} \bOmega d \tau' $, $ \frac 14 L^2 \bOmega + \varkappa  \int_0^{\tau} \bOmega d \tau' $. We will see below that the important contribution comes from exponentially large terms: each component of $\bOmega$ either grows or decreases exponentially. If it grows, for time large enough one can neglect the second terms in the sums (since $r_d \ll l \ll L$, $\int \bOmega d\tau \sim \lambda_3^{-1} \bOmega$). If it decreases, after some time the first term of each sum becomes smaller than the second, so one cannot neglect the viscous term.  However, in the case one can substitute a constant for $\int_0^{\tau} \bOmega d \tau' $. So, in any case the integral in the exponent can be replaced by a constant:
\begin{equation} \label{13-5}
\frac 14 l^2 \bOmega + \varkappa  \int_0^{\tau} \bOmega d \tau'  \simeq \frac 14 l^2 \bOmega + \varkappa /\lambda_3 {\bf C}
\end{equation}
where $\bf C$ is a constant matrix with elements $\sim 1$.
This simplification makes $\Psi$ a local functional of $\bOmega$. Thus,
 %(i.e., the functional that depends only on momentarily values of $\rho(\tau)$
 % and not on their integrals or derivatives). So,
 the functional average in (\ref{beta-average})
can be reduced to an ordinary multiple integral:
$$
%\begin{array}{l}
\beta_0 =
 \int d \tau \left \langle   \Psi (\bOmega (\tau)) \right \rangle_{\bf A}
 = \int d \tau  \int d\bOmega P(\tau, \bOmega)  \Psi(\bOmega)
%\end{array}
$$
where  $P$ is the probability density.

% Now $\tau$ and $\rho_i$ become independent variables.
%  %in particular, one can change the order of integration.

\section{ Properties of the evolution matrix}
It is convenient to  make the Iwasawa decomposition for the evolution matrix:
\begin{equation} \label{Iwasawa}
{\bf W} = {\bf  z d R } \  ,
\end{equation}
$$
% R \in SO(3) , \quad
 {\bf d} = \mbox{diag} \{e^{\rho_i } \}     \ , \quad           %\quad z_{i>j}=0 , \  z_{ii}=1
$$
where $\bf z$ is an upper triangular matrix with unities at diagonals, $\bf R$ is a rotation matrix, $\bf d$ is a diagonal matrix.
The incompressibility condition implies that
\begin{equation} \label{sum-rho}
\rho_1 + \rho_2 + \rho_3 =0
\end{equation}
According to (\ref{evolution_W}),(\ref{Iwasawa}), the stochastic processes $\rho_i$,$\bf z$ and $\bf R$ are functionals of ${\bf A}(t)$.
The long-time asymptotic behavior of these three components
is known to be quite different  \cite{Let-survey}: as $t \to \infty$,  ${\bf z}(t)$ stabilizes with unitary probability at some random value that depends on the realization of the process;  $\bR(t)$ remains changing randomly, and $\rho_i(t)/t$ converge (with unitary probability) to finite non-random limits $\lambda_i$  (Lyapunov indices):
$$ %\begin{equation} \label{lambdadef}
\lambda_i =  \lim \limits_{t\to \infty} \frac{\rho_i}{t},  \qquad \lambda_1 \le \lambda_2 \le \lambda_3
$$ %\end{equation}
%These indices are not random, they depend on statistics of ${\bf A}(t)$ but not on its realization.

In \cite{JOSS1,JOSS2} statistical properties of $\rho_i$ and $\bf z$ for large but finite time
%, in particular the Lyapunov exponents,
are expressed in terms of the statistical properties of $\bf A$.
The variables $\rho$ and $z$ are not independent; from (\ref{evolution_W}) 
%and (\ref{Iwasawa}) one can derive  relations between
it follows that, 
 %The isotropy of $\bf A$ imposes rigorous constraints on  the statistics of $\rho_i$ and $\bf z$. In particular, 
 to logarithmic accuracy,  the non-trivial components of $\bf z$ can be expressed as functions of $\rho_i$ at the same time and for the same realization \cite{JOSS1}. The matrices $\bf R$ do not contribute to $\bOmega$; so, with account of (\ref{sum-rho}),
 average over the process ${\bf A}(t)$ is equivalent to the average over  $\rho_1(t)$, $\rho_2(t)$ (with appropriate weight). The probability density of $\rho_1,\rho_2$ can be  expressed in terms of  cumulant function \cite{Klyackin,klassiki-large-dev,JOSS1}:
\begin{equation} \label{f}
f(\rho_1, \rho_2, \tau) = \int_{-i\infty}^{+i\infty} d \sigma_1 d\sigma_2  e^{- \sigma_1 \rho_1 - \sigma_2 \rho_2  + \tau W(\sigma_1, \sigma_2)}
\end{equation}
 The cumulant function $W$ is defined by $e^{W(\sigma_1, \sigma_2)}= \left \langle e^{\sigma_1 \rho_1+\sigma_2 \rho_2} \right \rangle$; in particular, it follows that Lyapunov exponents can be expressed as $\lambda_{1,2}= \left. \d W/\d \sigma_{1,2} \right|_{{\mathbf \sigma}=0}$.
For real arguments $\sigma_j$, $W$ is a real concave function with negative minimum, and $W(0,0)=0$.
 In \cite{JOSS1, PREnew} the procedure of derivation $W(\sigma_1,\sigma_2)$ from the cumulant function of the random process $A(t)$ is described in details. The result is
\begin{equation} \label{X-Arelation}
W (\sigma_1, \sigma_2) = w_A (\sigma_1-1,\sigma_2,1) -  w_A (-1,0,1)
\end{equation}
where $w_A(\boldsymbol{\eta} )$ is the 'diagonal part' of the cumulant function of $A(t)$,
$
w_A(\eta_1,\eta_2,\eta_3) = w_A \left. \left(  \bfeta  \right) \right|_{\eta_{i\ne j}=0}
$.

The condition of statistical isotropy of the process ${\bf A}(t)$ together with the incompressibility condition
$
\sum _{j=1..3} \frac {\d w_A}{\d \eta_j} = 0
$
make rigorous constraints on $w_A$ \cite{PREnew}: it can be reduced to a function of two variables,
\begin{equation} \label{w-ab}
w_A(\boldsymbol{\eta} )  = \tilde{w} (\alpha , \beta) \ ,
\end{equation}
$$ \textstyle
\begin{array}{c}
\alpha = \sum{\eta_j^2} - \frac 13 \left( \sum{\eta_j} \right)^2  , \ \\
  \beta =\sum{\eta_j} \cdot \sum{\eta_j^2} - \frac 29
\left( \sum{\eta_j} \right)^3 - \left( \sum{\eta_j^3} \right)
\end{array}
$$
According to (\ref{w-ab}),(\ref{X-Arelation}), all the pairs $(\sigma_1,\sigma_2)$ that correspond to the same $(\alpha,\beta)$ produce the same $W$.
 In particular, the set $\boldsymbol{\eta} _0=(-1,0,1)$ corresponds to $\alpha=2, \beta=0$; so, $W=0$ for all the points in the $\sigma$-plane that correspond to $(\alpha,\beta)=(2,0)$, including the   six 'universal' points with integer coordinates: $(0,0),(2,1),(3,0),(2,-2),(0,-3),(-2,-1)$.
The minimum of $W$ is achieved at the point $(\sigma_1, \sigma_2)=(2,1)$ which corresponds to $\alpha=\beta=0$ The dependence of $W$ on $\sigma_1, \sigma_2$ is illustrated in Fig.~1.

\section{ Map of $\Psi(\rho_1,\rho_2)$}
Substituting (\ref{Iwasawa}) into (\ref{psi-polu}) and making use of (\ref{13-5}), we get
\begin{equation}\label{E:Psi_Tr}
\Psi  =
 \frac{4 \pi  \eps_B}{3} \frac{ \left[ \Bigl(\mathrm{Tr} \bigl(\mathbf{z}\,\mathbf{d}^2\,\mathbf{z}^T\bigr)\mathrm{Tr} \bigl(\mathbf{d}^2\,\mathbf{g}\bigr)-
\mathrm{Tr} \bigl(\mathbf{z}\,\mathbf{d}^2\,\mathbf{g}\,\mathbf{d}^2\,\mathbf{z}^T\bigr)\Bigr) \right.}
 {\sqrt{\mathrm{det}\mathbf{G \, G}_L}} \ ,
\nonumber
\end{equation}
where
\begin{equation}    \nonumber %\label{Gg}
\begin{array}{l}
\mathbf{G}  {\simeq}\mathbf{d}^2 +\left(\frac{r_d}{l}\right)^2\,\mathbf{ z}^{-1} \mathbf{C}\,{\left(\mathbf{ z}^{-1}\right)}^{T};
\\
\mathbf{G}_L {\simeq}\mathbf{d}^2 +\left(\frac{r_d}{L}\right)^2\,\mathbf{ z}^{-1} \mathbf{C}\,{\left(\mathbf{ z}^{-1}\right)}^{T}
\end{array}
\end{equation}
With account of asymptotic dependence $\bf z$ on $\rho_i$, $\Psi$ is a univocal function of $\rho_1,\rho_2$. One can show~\cite{PREnew} that its maximum is situated in the ray
\begin{equation} \label{maximum-psi}
\rho_1< - R \equiv - \log L/r_d,\,\,\,\,\,\, \rho_2 = - r \equiv - \log l/r_d
\end{equation}
(see Fig.2), and is equal to
$$
\Psi_{\mathrm{max}} \simeq\varepsilon_B\frac{L\,l^3}{r_d^4}.
$$
Near the ray the function decreases exponentially; in Fig.2 the level line corresponding to $\Psi=\Psi_{max} r_d/l \ll \Psi_{max}$ is shown to illustrate this rapid decrease. In the vicinity of the maximum, $\Psi$ can be approximated by
\begin{equation} \label{Psi-near-max}
\frac{\Psi}{\Psi_{\mathrm{max}}}=\left[ \begin{array}{lll}
{\mathrm e}^{(\rho_2+r)},& \rho_2<-r,\, \rho_1<-R &  (I)
\\
\mathrm e^{-2(\rho_2+r)},& \rho_2>-r, \, \rho_1<-R &  (II)
\\
\mathrm e^{-(\rho_1+R)-2(\rho_2+r)},&\rho_2>-r, \, \rho_1>-R &  (III)
\\
\mathrm e^{-(\rho_1+R)+(\rho_2+r)},&\rho_2<-r, \, \rho_1>-R &  (IV)
\end{array} \right.
\end{equation}

\begin{figure}[t]
%\vspace*{-0.37cm}
\hspace*{-0.5cm}
\includegraphics[width=8.5cm]{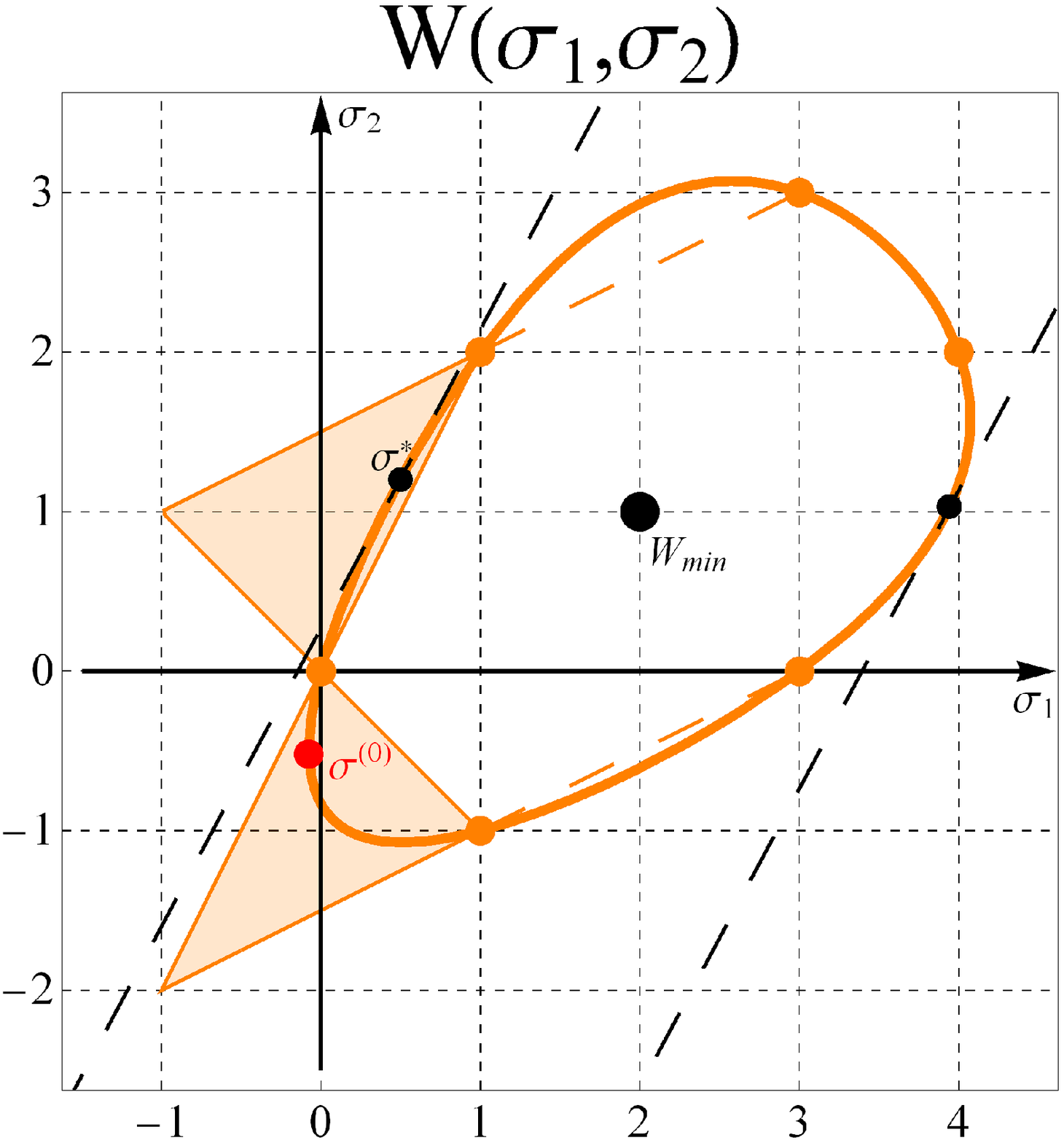}
%\vspace*{-0.9cm}
%\includegraphics[width=16.5cm]{points.png}
\caption{
Illustration for the level line W = 0 in the $(\sigma_1,\sigma_2)$ plane (the case $\lambda_2>0$), six universal
points and the point of minimum of $W$ are presented. For some given $\rho_1,\rho_2$, the left of the two parallel tangent straight lines with the slope $(\rho_2, -\rho_1)$ picks out the point $\sigma^*$ defined by (\ref{E:saddle_conditions_2}).
The point  with minimal abscissa on the level line corresponds to  the zero-approximation to $\sigma^*$ for $\rho_1,\rho_2$ satisfying (\ref{maximum-psi}).
}
\end{figure}

\begin{figure}[b]
%\vspace*{-0.37cm}
%\hspace*{-0.5cm}
\includegraphics[width=8.5cm]{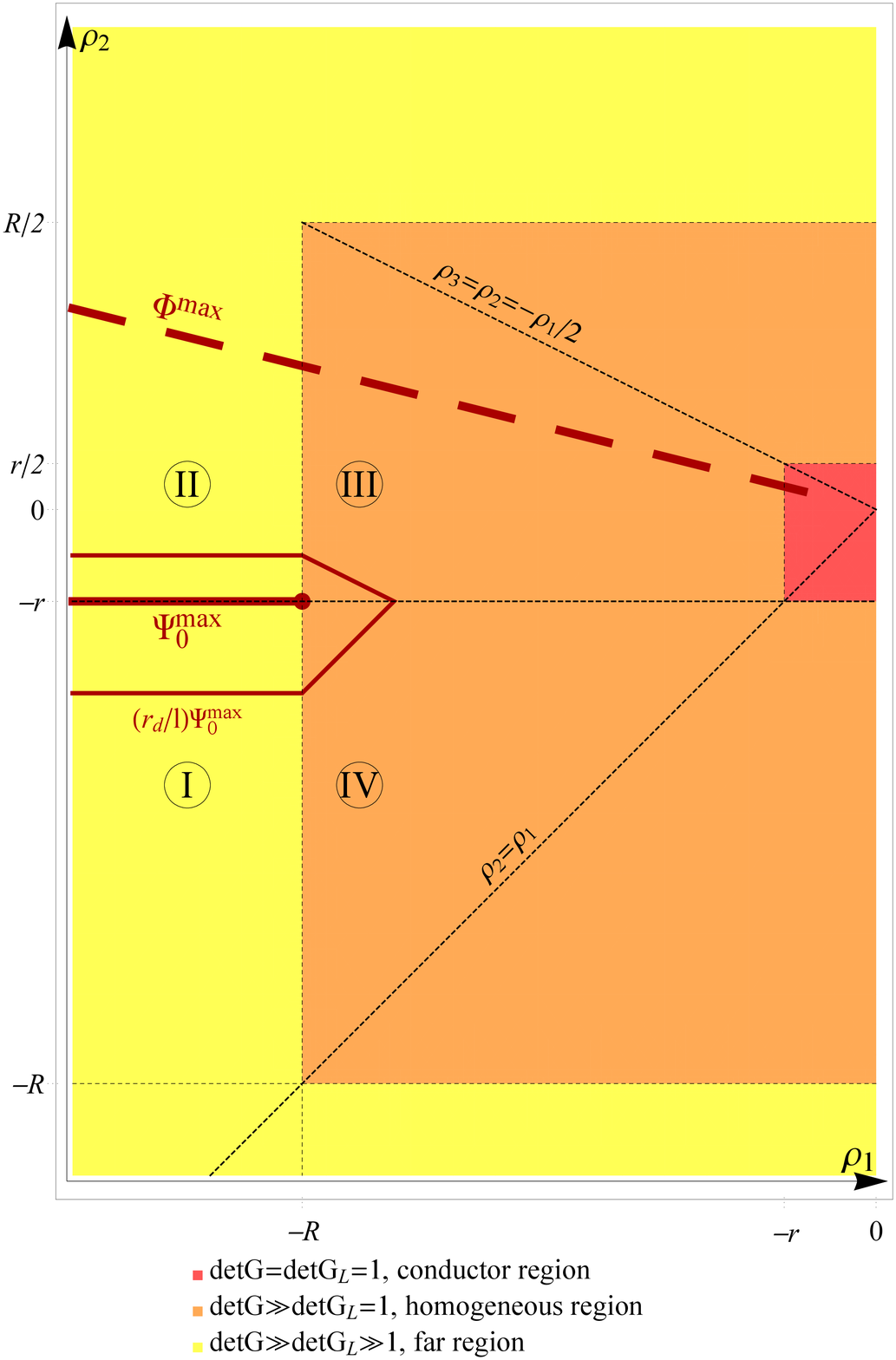}
%\vspace*{-0.9cm}
%\includegraphics[width=16.5cm]{points.png}
\caption{
The  $\rho_1$, $\rho_2$ plane with lines of maximal $\Psi$ (thick solid) and $\Phi$ (thick dashed). One level line of $\Psi$ is shown. The position of the dashed line corresponds to the case $\lambda_2>0$. The inner rectangular region of the plane corresponds to the ideal conductor, in the outer region the influence of finite pumping scale $L<\infty$ must be taken into account. Thin dashed lines demarcate regions with different approximations for $\Psi$.
}
%\label{F:Lera2}
\end{figure}

\section{ Time integral of probability density}
The mean square magnetic field can now be rewritten as
$$  %\begin{equation}\label{E:beta_integral}
\beta_0 = \int  \mathrm{d}\rho_1 \mathrm{d}\rho_2 \Phi(\rho_1,\rho_2) \Psi(\rho_1,\rho_2) , \ \
\Phi = \int d \tau f(\rho_1,\rho_2,\tau)
$$
With account of (\ref{f}) and making use of the saddle-point method ($\rho=\sqrt{\rho_1^2+\rho_2^2}\gg1$) we get
\begin{equation} \label{Saddle}
\Phi \simeq \frac 1{|\lambda_1|} \mathrm{e}^{\textstyle \tau^* W(\sigma_1^*,\sigma_2^*)-\sigma_1^*\rho_1-\sigma_2^*\rho_2},
\end{equation}
where the saddle values $\sigma_1^*, \sigma_2^*, \tau^*$ are determined by
$$ %\begin{equation}\label{E:saddle_conditions}
\left\{\begin{array}{l}
W(\sigma_1^*,\sigma_2^*)=0
\\
\tau^* \frac{\partial}{\partial \sigma_1^*}W(\sigma_1^*,\sigma_2^*)-\rho_1=0
\\
\tau^* \frac{\partial}{\partial \sigma_2^*}W(\sigma_1^*,\sigma_2^*)-\rho_2=0.
\end{array}\right.
$$ %\end{equation}
Fig. 1 proves that this system  has a unique solution for any possible $W$. Actually,
excluding $\tau^*$ we get
\begin{equation}\label{E:saddle_conditions_2}
\left\{\begin{array}{l}
W(\sigma_1^*,\sigma_2^*)=0
\\
 \rho_2\frac{\partial}{\partial \sigma_1^*}W(\sigma_1^*,\sigma_2^*) -\rho_1 \frac{\partial}{\partial \sigma_2^*}W(\sigma_1^*,\sigma_2^*)=0
\end{array}\right.
\end{equation}
with additional condition $\frac{\partial}{\partial \sigma_2^*}W(\sigma_1^*,\sigma_2^*)/\rho_2 >0$ responsible for positive $\tau^*$.  The first equation corresponds to the level line $W=0$ in the $\sigma_1,\sigma_2$-plane; the second solution picks out the points where the tangent to the level line is parallel to the direction $(\rho_2, -\rho_1)$. From concavity of the level line it follows that there are two such points, and exactly one of them corresponds to $\tau^*>0$.

Returning to the $\rho_1,\rho_2$ plane and considering polar coordinates, it is natural to await maximal values of $\Phi$ (for given $\rho$) along the direction $(\rho_1,\rho_2)\propto (\lambda_1,\lambda_2)$ (see Fig.2).$^1$\footnotetext[1]{
Formal proof of the fact  can be found by expressing probability density in terms of the Kramer function,
$f(\rho_1,\rho_2,t)=\mathrm{e}^{-S(\rho_1/t-\lambda_1,\rho_2/t-\lambda_2)t}$; $S(x,y)$ has the minimum at $x=y=0$.}%%%%%%%%%%%%
 \, Taking into account that $\d W /\d{\sigma_1}(0,0)=\lambda_1$, $\d W / \d{\sigma_2}(0,0)=\lambda_2$, one can easily check that the pair $\sigma_1^*=\sigma_2^*=0$  satisfies the conditions (\ref{E:saddle_conditions_2}) for all $\rho_1,\rho_2$ in this direction; so, from (\ref{Saddle}) one derives that  $\Phi$ does not depend on $\rho$ for all the points of the ray,
$$ %\begin{equation}\label{E:Phi_max}
\Phi_\mathrm{max}\left(\rho_1, \rho_2\right)=\Phi\Bigl(%\overbrace
{\frac{\lambda_1}{\sqrt{\lambda_1^2+\lambda_2^2}}} %^{\cos\phi_0}
\,\rho, %\overbrace
{\frac{\lambda_2}{\sqrt{\lambda_1^2+\lambda_2^2}}}%^{\sin\phi_0}
\,\rho\Bigr)=\mathrm{Const}.
$$ %\end{equation}
In all other directions $\Phi$ decreases exponentially at large $\rho$.

We now concentrate on the behavior of $\Phi$ in the vicinity of the maximum of $\Psi$, i.e., the ray (\ref{maximum-psi}).  Since $|\rho_2| \ll |\rho_1|$, one can use the method of successive approximations with small parameter $\rho_2$. Substituting
$$
\sigma_1^*=\sigma_1^{(0)}+\rho_2 \sigma_1^{(1)},\,\,\,\,\,\sigma_2^*=\sigma_2^{(0)}+\rho_2 \sigma_2^{(1)}
$$
in (\ref{E:saddle_conditions_2}) and expanding into a series in $\rho_2$, we get to the zeroth order:
$$
W^{(0)}=0  \ , \quad
 \frac{\partial}{\partial \sigma_2^{(0)}}W^{(0)}=0
 $$
The second condition corresponds to the vertical tangent to the level line in Fig.1. The additional condition $\tau^*>0$ singles out the left of the two points.

Expanding the first equation in (\ref{E:saddle_conditions_2}) to the first order, we find
$$ %\begin{equation}\label{E:sigma_ints}
\sigma_1^{(1)}\frac{\partial}{\partial \sigma_1^{(0)}}W^{(0)}+\sigma_2^{(1)}\overbrace{\frac{\partial}{\partial \sigma_2^{(0)}}W^{(0)}}^{=0}=0\Rightarrow\sigma_1^{(1)}=0.
$$ %\end{equation}
From (\ref{Saddle}) we then obtain
\begin{equation} \label{E:Phi_saddle}
\Phi\propto\,\mathrm{e}^{-\sigma_1^{(0)}\rho_1-\sigma_1^{(1)}\rho_1-\sigma_2^{(0)}\rho_2} = \mathrm{e}^{-\sigma_1^{(0)}\rho_1-\sigma_2^{(0)}\rho_2}
\end{equation}
The 'universal' points definitely belonging to the curve $W=0$ together with its concavity oblige the left part of the curve to belong to the two highlighted  triangles  in Fig.1. This restricts the coordinates $\sigma_1^{(0)}$,  $\sigma_2^{(0)}$ by the boundaries of these triangles:
\begin{equation} \label{granicy}
\begin{array}{c}
 0\geq\sigma_1^{(0)}>-1 \ ,\quad 3/2>\sigma_2^{(0)}>-2 \ ,\\
 %\quad
 \mathrm{sign}\,\sigma_2^{(0)}=-\mathrm{sign}\,\lambda_2
 \end{array}
\end{equation}
Thus, if $\sigma_1^{(0)} \ne 0$ (which is, $\lambda_2 \ne 0$),  the exponent in (\ref{E:Phi_saddle}) is negative for any $-\rho_1$ large enough.
%In vertical direction, the function $\Phi$ increases in one direction and decreases... - ��� ���������� � ����! ���� $sign \rho \ne sign (\rho+r)$

\section{ Convergence and calculation of $\beta_0$ }
The function $\Psi$  is more concentrated near its maximum than $\Phi$, i.e.,  decreases faster as the point moves away from it. So, the main contribution to $\beta_0$ comes from the region of maximal $\Psi$.
%One can check that in other directions, in particular in the direction of maximal $\Phi$, the contribution
%to the integral is not essential.
Actually, the integrand $\Psi \Phi$ can be easily calculated in the vicinity of the maximum by multiplication (\ref{Saddle}) and (\ref{Psi-near-max}), and one can check that the maximum is situated at $(\rho_1,\rho_2)=(-R,-r)$ and coincides with the maximum of $\Psi$ if $\sigma_2^{(0)}<1$. We now  formally extrapolate these formulas to the whole plane $(\rho_1,\rho_2)$,
since the difference is essential only in the regions where the contribution to the integral is negligible:
$$
\begin{array}{r}
\frac{\beta_0 }{\Psi_{max} \Phi_{max}} = \left(  \int \limits_{-\infty}^{-R} e^{-\sigma_1^{(0)} \rho_1} d\rho_1  +
  \int \limits_{-R}^{\infty} e^{-(\sigma_1^{(0)}+1) \rho_1 -R} d\rho_1  \right) \\
\times   \left(  \int \limits_{-\infty}^{-r} e^{(-\sigma_2^{(0)}+1) \rho_2 +r} d\rho_2  +
  \int \limits_{-r}^{\infty} e^{-(\sigma_2^{(0)}+2) \rho_2 -2r} d\rho_2  \right)
\end{array}
$$
With account of (\ref{granicy}), we see that all the integrals converge if $\sigma_1^{(0)} \ne 0$, $\sigma_2^{(0)} <1$.
The first of these conditions excludes the case $\lambda_2=0$: in time-symmetric velocity gradient field the mean-square of magnetic field diverges linearly. The second condition prohibits the cases corresponding to   large negative $\lambda_2$. In these cases the point of maximum of $\Phi\Psi$  in the $(\rho_1,\rho_2)$ plane does not coincide with the maximum of $\Psi$, and our decomposition is not valid. One should consider them more accurately. Here we are not interested in them, since the case $\lambda_2<0$ does not correspond to physical reality \cite{lambda2=1/4}.

So, for $-2<\sigma_2^{(0)} <1$ we take the integrals, with account of (\ref{maximum-psi}) and (\ref{Psi-near-max}) we obtain:
  \begin{equation} \label{general-itog}
\beta_0   \simeq \frac{\varepsilon_B}{|\lambda_1|} \frac{Ll^3}{r_d^4}
 K^{-1}
  \left( \frac L{r_d}\right)^{\sigma_1^{(0)}}
  \left( \frac l{r_d} \right)^{\sigma_2^{(0)}}
\end{equation}
$$
K = \frac 13 { (-\sigma_1^{(0)})(\sigma_1^{(0)}+1)}
{(1-\sigma_2^{(0)})(2+\sigma_2^{(0)})}
$$

\section{ Small $|\lambda_2|$}
We now calculate $\beta_0$ for velocity gradient statistics close to time reversible \cite{IZ}: $|\lambda_2|\ll|\lambda_1|$.  In this case the point $(\sigma_1^{(0)},\sigma_2^{(0)})$ is close to $(0,0)$.
We expand  $\sigma_1^{(0)},\sigma_2^{(0)}$ into a series to the second order of $\lambda_2$:
$$ %\begin{equation}
\begin{array}{l}
\left\{\begin{array}{l}
\sigma_1^{(0)}=\xi_1\lambda_2+\zeta_1\lambda_2^2
\\
 \sigma_2^{(0)}=\xi_2\lambda_2+\zeta_2\lambda_2^2
\end{array}\right.\Rightarrow  \\
\left\{
\begin{array}{l}
\xi_1=0
\\
\lambda_2+D_2\xi_2\lambda_2=0
\\
\lambda_1\zeta_1\lambda_2^2+\xi_2\lambda_2^2+\frac{1}{2}D_2\xi_2^2\lambda_2^2=0
\end{array}
\right.
\end{array}
$$ %\end{equation}
Here $D_2$ is the dispersion, $D_2\propto |\lambda_1|$. Thus,
$$
\xi_1=0  \ , \quad
\xi_2=-\frac{1}{D_2} \ , \quad \zeta_1=-\frac{1}{2|\lambda_1|D_2}
$$
and, to the main non-vanishing order,
$$
\sigma_1^{(0)} = -\frac{\lambda_2^2}{2D_{2}|\lambda_1|} \ , \quad
\sigma_2^{(0)} = -\frac{\lambda_2}{D_{2}}
$$
Substituting this into (\ref{general-itog}), we get
\begin{equation}%\boxed{\label{E:beta_fin}
\label{small-l2-itog}
\beta_0 \propto\, \varepsilon_B \frac{  %|\lambda_1|
D_{2}}{\lambda_2^2}\left(\frac{L}{r_d}\right)^{1-\frac{\lambda_2^2}{2D_{2}|\lambda_1|}}
\left(\frac{l}{r_d}\right)^{3-\frac{\lambda_2}{D_{2}}}
\end{equation}
One can compare this with the two-point correlation function obtained in \cite{PREnew} for the range $r_d \ll r \ll l$:
$$ %\begin{equation}
\beta(r)\propto\mathrm{min}\left(\frac{\lambda_3 D_{2}}{\lambda_2^2},\log\frac{L}{l}\right)\left(\frac{l}{r_d}\right)^{1-\frac{\lambda_2^2}{2D_{2}\lambda_3}}\left(\frac{l}{r}\right)^{3-\frac{\lambda_2}{D_{2}}}.
$$ % \end{equation}
One can see that if ${\lambda_3 D_{2}}/{\lambda_2^2}<\log{L}/{l}$, $\beta(r)$ does not depend on the scale  $L$, and it follows $\beta(0)/\beta(r_d) \sim L/l$.
% ^{1-\lambda_2^2/2D_{2}}$.
 As a function of $\lambda_2$, the ratio is almost constant; $\beta_0$ and $\beta(r)$ both increase in proportion to $\lambda_2^{-2}$, and the shape of $\beta(r)$ near its maximum does not change significantly while $\lambda_2^2 >\lambda_2^{*2} = \lambda_3^2 \log{L}/{l}$. However, as $\lambda_2$ becomes smaller than $\lambda_2^*$, the $\beta(r)$ dependence  of $\lambda_2$ freezes outside the radius $\sim r_d$, and $\beta(r_d) \sim \log{(L/l) } \, (l/r_d)^4$ does not depend on $\lambda_2$ any more. To the contrary, $\beta_0$ remains behaving  as $\lambda_2^{-2}$; so, for these smallest $\lambda_2$ the shape of the $\beta(r)$ curve near the maximum transforms essentially, the ratio $\beta(0)/\beta(r_d)$ becoming infinite as $\lambda_2=0$. This is, in particular, what we get in the Gaussian velocity field.

\section{ No-feedback condition: estimate of pressure and energy}
We now estimate the possibility of magnetic field feedback on velocity dynamics.
The divergence of the Navier-Stokes equation with account of the Lorentz force is
$$
\nabla \cdot (\bv \nabla) \bv = - \Delta p + \frac 1{4\pi} \nabla \cdot (\bB \nabla) \bB
$$
In the viscous range, the velocity term is negligible, and the pressure can be found from this equation by
$$  %\begin{equation}
p = p_0 + \int \frac{\nabla \cdot (\bB \nabla) \bB}{r} d\br
$$  % \end{equation}
Here $p_0 \sim \left \langle v^2 (r_{\eta}) \right \rangle$ is the pressure component produced by velocity field at scales larger than $r_{\eta}$.

The region $L$ of magnetic field generation is small compared to the Kolmogorov viscous scale. Thus, at the scale $r_{\eta}$ one can expand the equation into multipoles:
$$
\begin{array}{rl}
p \simeq & p_0 + \frac 1R  \int _L \nabla \cdot (\bB \nabla) \bB d\br + \left( \int _L \nabla \cdot (\bB \nabla) \bB  \br d\br \right) \cdot \frac {\bf R}{R^3}  \\
+& \left( \int _L \nabla \cdot (\bB \nabla) \bB  r_i r_j d\br \right) \frac {R_i R_j}{R^5} + \dots
\end{array}
$$
where $R\sim r_{\eta} \gg L$. The integrals in the first two summands are equal to zero, and for the quadrupole moment we get
$$
p(r_{\eta}) =  \left( 2 \int B_i B_j d\br \right)  \frac {R_i R_j}{R^5} \sim \left \langle B^2 \right \rangle \frac {L^3}{r_{\eta}^3}
$$
The absence of feedback implies that the pressure generated by Lorentz force is much smaller than $p_0$, i.e.,
\begin{equation}
\label{inequ}
%\left \langle B^2 \right \rangle 
\beta_0 \frac {L^3}{r_{\eta}^3} < \left \langle v^2 (r_{\eta}) \right \rangle
\end{equation}
One can see that this condition coincides with the requirement that magnetic energy averaged over the volume
 $\sim r_{\eta}^3$ is smaller than the average kinetic energy.

Eq.(\ref{inequ}) provides an upper limit for the  %restriction on maximal 
pumping power possible without the feedback. However, both $\phi$ and $\eps_B$ depend on the diffusivity; the 'source' current ${\bf j}_S (\br,t)$ is the independent parameter that governs the driving force,
% The driving force $\phi$ is produced by some 'source' current ${\bf j}_S (\br,t)$:
 $$
 \phi = \frac {4 \pi \varkappa}{c} \mathbf{rot} {\bf j}_S
 $$
The spatial correlation scale of ${\bf j}_S$ is $l$; from (\ref{phi-kvadrat}) it follows
$$
\frac {\left \langle j_S^2 \right \rangle}{c^2} \sim
\frac{\left \langle \left( \mathbf{rot} {\bf j}_S \right)^2 \right \rangle}{c^2} l^2 \sim \frac 1{\varkappa^2} \left \langle \phi^2 \right \rangle l^2 \sim
 \frac 1{\varkappa^2}  \frac{\eps_B}{\tau_{\phi}} l^2
$$
where $\tau_{\phi}$ is the correlation time for $\phi$ and ${\bf j}_S$. Substituting this in (\ref{small-l2-itog}) and then in  (\ref{inequ}),  we can compare $\left \langle {\bf j}_S ^2  \right \rangle$ to the characteristic vorticity of the flow. The condition of feedback absence then reads as
\begin{equation}
\label{fin-est}
\frac {\left \langle {\bf j}_S^2 \right \rangle}{c^2 \left \langle  \bomega^2 \right \rangle} <
\left( \frac{\lambda_2}{\lambda_1} \right)^2
 \frac{1}{\tau_{\phi} \lambda_1}      \frac{r_{\eta}^5}{l L^4}
\end{equation}
We see that even for $L\sim r_{\eta}$, which is the upper limit of the range under consideration, 
%in (\ref{inequalities}),
the right-hand side of the inequality is still much larger than unity. So, a wide range of scale relations allows the stationary solution without feedback. We also note that the estimate (\ref{fin-est}) does not depend on  the magnetic diffusivity  (or the diffusivity scale).  This is a consequence of a combination of two effects: on one hand, in well-conducting fluid   the energy flux ($\eps_B$)
produced by given current is small ($\sim \varkappa^2$); on the other hand, the amplitude of magnetic field fluctuations is larger in the medium with smaller magnetic diffusion (\ref{small-l2-itog}).

%There is no feedback if mean-square magnetic field is much less than mean-square velocity averaged over the %Kolmogorov scale $\eta$. According to the Kolmogorov's 2/3 law, with account of $v\sim \eta \cdot \lambda_1$, %we get
%$$
%\langle v^2 \rangle \sim \varepsilon^{2/3} \eta^{2/3} \sim  \frac{\varepsilon}{\lambda_1}
%$$
% A very rough upper estimate for magnetic energy inside one 'blob' of the size $L$ can be found by %multiplying (\ref{small-l2-itog}) by $L^3$, then average over $\eta$ is
% $ \langle B^2 \rangle \sim \beta_0 \frac {L^3}{\eta^3} $.
% Thus, the condition $\langle B^2 \rangle \ll \langle v^2 \rangle $ is equivalent to
% $$
% \frac {\varepsilon_B}{\varepsilon} \ll \left( \frac{\lambda_2}{\lambda_1} \right)^2
% \left( \frac{r_d}{L} \right)^4    \left( \frac{\eta}{l} \right)^3
% $$
%We see that even for $\varepsilon_B = \varepsilon$ there still exists a range of scale  relations that allows %the stationary solution without feedback.

\section{ Conclusion}
Summarizing, we find stationary solution for forced magnetic field embedded in viscous turbulent flow.
In the solution, the increase of magnetic field energy density is stopped as a result of combination of two effects:  finite size  of the pumping region and  non-Gaussianity of velocity gradient statistics. The resulting mean-square magnetic field is presented in (\ref{general-itog}) and, for small deviation from Gauss statistics, in  (\ref{small-l2-itog}).
The estimate of possible energies and driving current intensity  shows  that  such stabilization is possible without feedback on velocity statistics in a wide range of parameters.

\acknowledgments
The authors are grateful to Prof. A.V. Gurevich for his permanent attention to their work.
The work of AVK was supported by the RSF  grant 20-12-00047.

\end{document}